\documentclass[10pt,conference,letter]{IEEEtran}

\usepackage{graphicx,graphics,subfigure,wrapfig}
\usepackage{epsfig}
\usepackage{psfrag}
\usepackage[usenames]{color}
\usepackage{amsmath,dsfont,amssymb}
\usepackage{cite,url,fancybox,balance}
\usepackage{array,multirow}
\usepackage[active]{srcltx}
\usepackage{float}
\usepackage{bm}
\usepackage{cmll}
\usepackage{enumerate}

\newtheorem{theorem}{Theorem}
\newtheorem{lemma}{Lemma}

\newfont{\mbb}{msbm10 scaled 1100}

\definecolor{red}{RGB}{210,20,50}
\newcommand{\red}{\textcolor{red}}
\definecolor{lblue}{RGB}{30,160,240}
\newcommand{\lblue}{\textcolor{lblue}}
\definecolor{green}{RGB}{75,180,130}

\definecolor{blue}{RGB}{0,0,255}
\newcommand{\blue}{\textcolor{blue}}

\definecolor{magenta}{RGB}{255,0,255}
\newcommand{\magenta}{\textcolor{magenta}}
\definecolor{orange}{RGB}{255,128,0}
\newcommand{\orange}{\textcolor{orange}}
\definecolor{dgreen}{RGB}{0,151,0}

\def\ind{{\mathds{1}}}
\def\R{{\mathds{R}}} 

\def\b0{{\bf 0}}

\begin{document}

\title{On the geometry of wireless network multicast in 2-D}

\author{
\authorblockN{Mohit Thakur}
\authorblockA{Institute for communications engineering,\\
Technische Universit\"{a}t M\"{u}nchen,\\
80290, M\"{u}nchen, Germany.\\
Email: mohit.thakur@tum.de}
\and
\authorblockN{Nadia Fawaz}
\authorblockA{Technicolor Research Center, \\
Palo Alto, CA, USA.\\
Email: nfawaz@mit.edu}
\and
\authorblockN{Muriel M\'{e}dard}
\authorblockA{Research Laboratory for Electronics,\\
Massachusetts Institute of Technology,\\
Cambridge, MA, USA.\\
Email: medard@mit.edu}
}

\vspace{-6mm}
\maketitle
\vspace{-2mm}

\begin{abstract}
We provide a geometric solution to the problem of optimal relay
positioning to maximize the multicast rate for low-SNR networks. The
network we consider consists of a single source, multiple receivers
and the only intermediate and locatable node as the relay. We
construct network the hypergraph of the system nodes from the
underlying information theoretic model of low-SNR regime that
operates using superposition coding and FDMA in conjunction (which
we call the ``achievable hypergraph model''). We make the following
contributions.
\begin{enumerate}
\item We show that the problem of optimal relay positioning maximizing the multicast rate can be completely
decoupled from the flow optimization by noticing and exploiting
geometric properties of multicast flow.
\item All the flow maximizing the multicast rate is sent over at most two paths, in succession. The relay position
depends on only one path (out of the two), irrespective of the
number of receiver nodes in the system. Subsequently, we propose
simple and efficient geometric algorithms to compute the optimal
relay position.
\item Finally, we show that in our model at the optimal relay position, the
difference between the maximized multicast rate and the cut-set
bound is minimum.
\end{enumerate}
We solve the problem for all $(P_{s},P_{r})$ pairs of source and
relay transmit powers and the path loss exponent $\alpha \geq 2$.

\end{abstract}

 \begin{keywords}
 Low-SNR, broadcast relay channel, geometry.
 \end{keywords}

\vspace{-2mm}
\section{INTRODUCTION} \label{sec:Introduction}
We primarily consider the problem of optimal relay positioning in
order to maximize the multicast rate in low-SNR networks consisting
of a single source $s$, a set of multiple receivers $T$ and an
arbitrarily locatable relay $r$, on a $2$-D Euclidean plane. In
\cite{Thakur-Fawaz-Medard-Infocom2011}, the authors previously
addressed this problem under a heavy and complex network flow
optimization framework. They showed that optimizing the relay
position can lead to a strong gain in the multicast rate.

In \cite{Thakur-Medard-Globecom2010} the authors introduced
equivalent hypergraph models for the low-SNR Broadcast (BC) and
Multiple Access  channels (MAC). The authors then derived an
achievable hypergraph model for the broadcast relay channel (BRC),
obtained by concatenating the equivalent BC and MAC hypergraphs.
This concatenated model follows from constraining the source and
relay to transmit using the optimal schemes for the low-SNR BC and
MAC: superposition coding and frequency division, respectively. In
this paper, building on this model, we solve geometrically the
problem of optimal relay positioning under the pretext of multicast
rate maximization, which is much simpler and efficient than the
solution proposed in \cite{Thakur-Fawaz-Medard-Infocom2011}.

Most importantly, we establish the fact that for a given low-SNR BRC
hypergraph $\mathcal{G(N,A})$, the multicast rate is maximized by
sending all the flow through at most two paths in succession,
independently of the number of destination nodes. This is a
consequence of simply maximizing the multicast min-cut. The
dependency of the multicast min-cut on the relay position is
essentially through a single path (out of the two), and this
motivates a simple geometric interpretation and  formulation of the
problem. It should be noted that, the ``optimal relay position''
refers to the position that maximizes the multicast rate over a
given achievable hypergraph, but in general the achievable
hypergraph model is not necessarily optimal in terms of meeting the
cut-set bound for low-SNR networks. On the other hand, the
achievable hypergraph model performs closely to the peaky binning
scheme in the case of a single destination
\cite{Fawaz-Medard-ISIT2010}, and enjoys an important practical
advantage of being easily scalable to more complicated topologies.
Finally, under our model the difference between the maximum
multicast rate and the cut-set bound is minimized at the optimal
relay position.

 In the proposed geometric approach, we decouple the problem of rate maximization from the problem of computing the optimal relay position.
This substantially reduces the complexity (compared to the flow
optimization based framework in
\cite{Thakur-Fawaz-Medard-Infocom2011})
 and also provides a great deal of insight in understanding the nature of such network planning problems.
Finally, we show that at the optimal position the difference between
the maximum multicast rate and the cut-set bound is minimized under
the achievable hypergraph model.

%

The paper is organized as follows. We introduce the low-SNR
achievable hypergraph model of the BRC in
section~\ref{sec:SysModel}. Then we prove certain geometric
properties of multicast in section~\ref{sec:Geometric}. The
computation of optimal relay position is divided in two parts,
section~\ref{sec:eq}
 for $P_{s}=P_{r}$ and section~\ref{sec:neq} for $P_{s} \neq P_{r}$.
 Finally, we conclude in section~\ref{sec:Conclusion}.

\begin{figure*}[tp]
\begin{center}
\psfrag{s}[cc][cc]{{\small $s$}} \psfrag{d1}[cc][cc]{{\small $t_1$}}
\psfrag{d2}[cc][cc]{{\small $t_2$}} \psfrag{0}[tc][tc]{{\small $0$}}
\psfrag{Ps}[cc][cc]{{\small $P$}} \psfrag{h1}[cc][cc]{{\small
$h_1$}} \psfrag{h2}[cc][cc]{{\small $h_2$}}
\psfrag{N1}[cc][cc]{{\small $N_0$}} \psfrag{N2}[cc][cc]{{\small
$N_0$}} \subfigure[BC]{
\includegraphics[width=0.24\columnwidth]{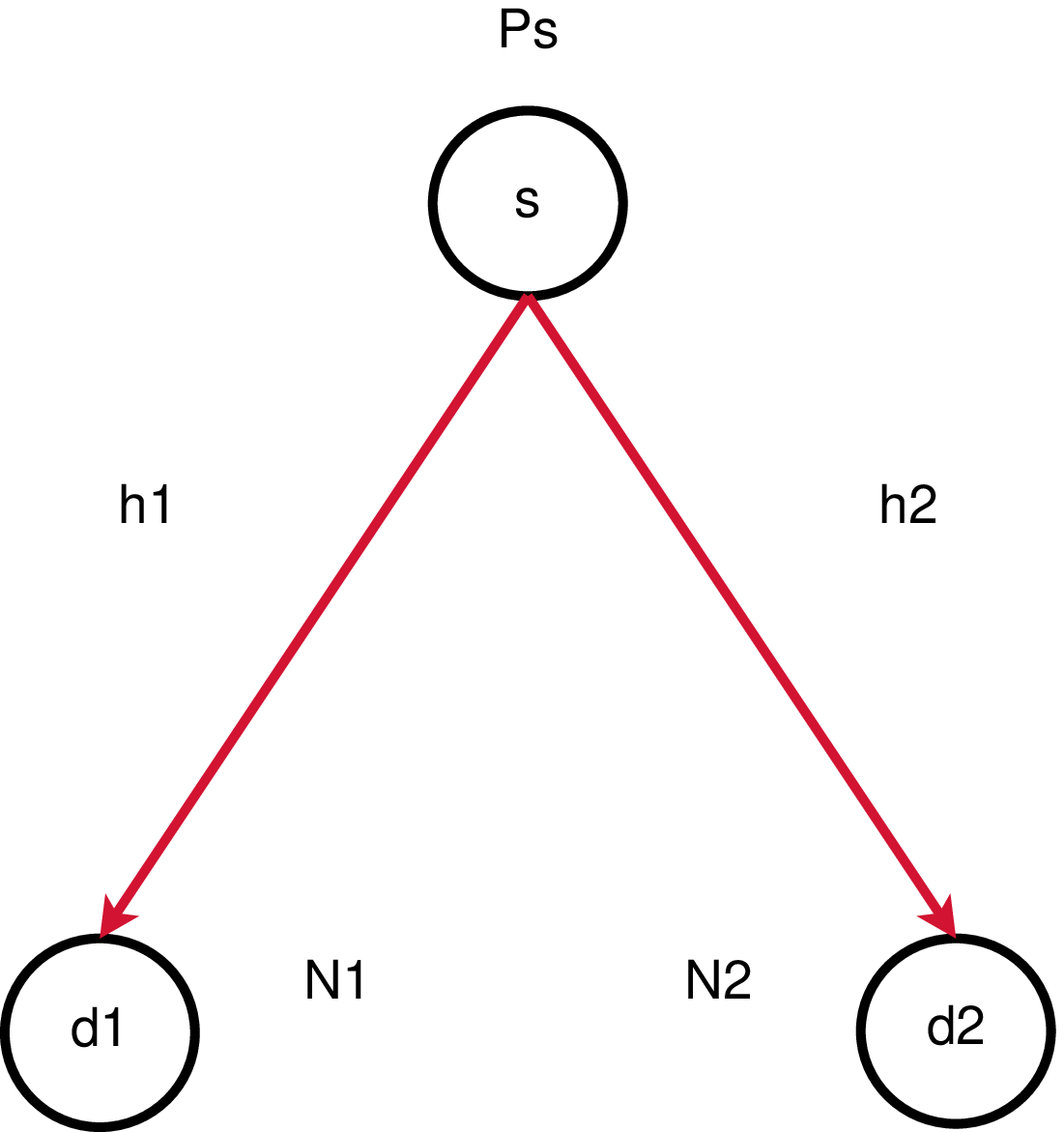}
\label{fig:BC} } \quad \psfrag{Cc}[cc][cc]{{\small \red{$R_c$}}}
\psfrag{Cb1}[cc][cc]{{\small \magenta{$R_1$}}}
\psfrag{Cb2}[cc][cc]{{\small \orange{$R_2$}}} \subfigure[Wideband BC
equivalent hypergraph]{
\includegraphics[width=0.26\columnwidth]{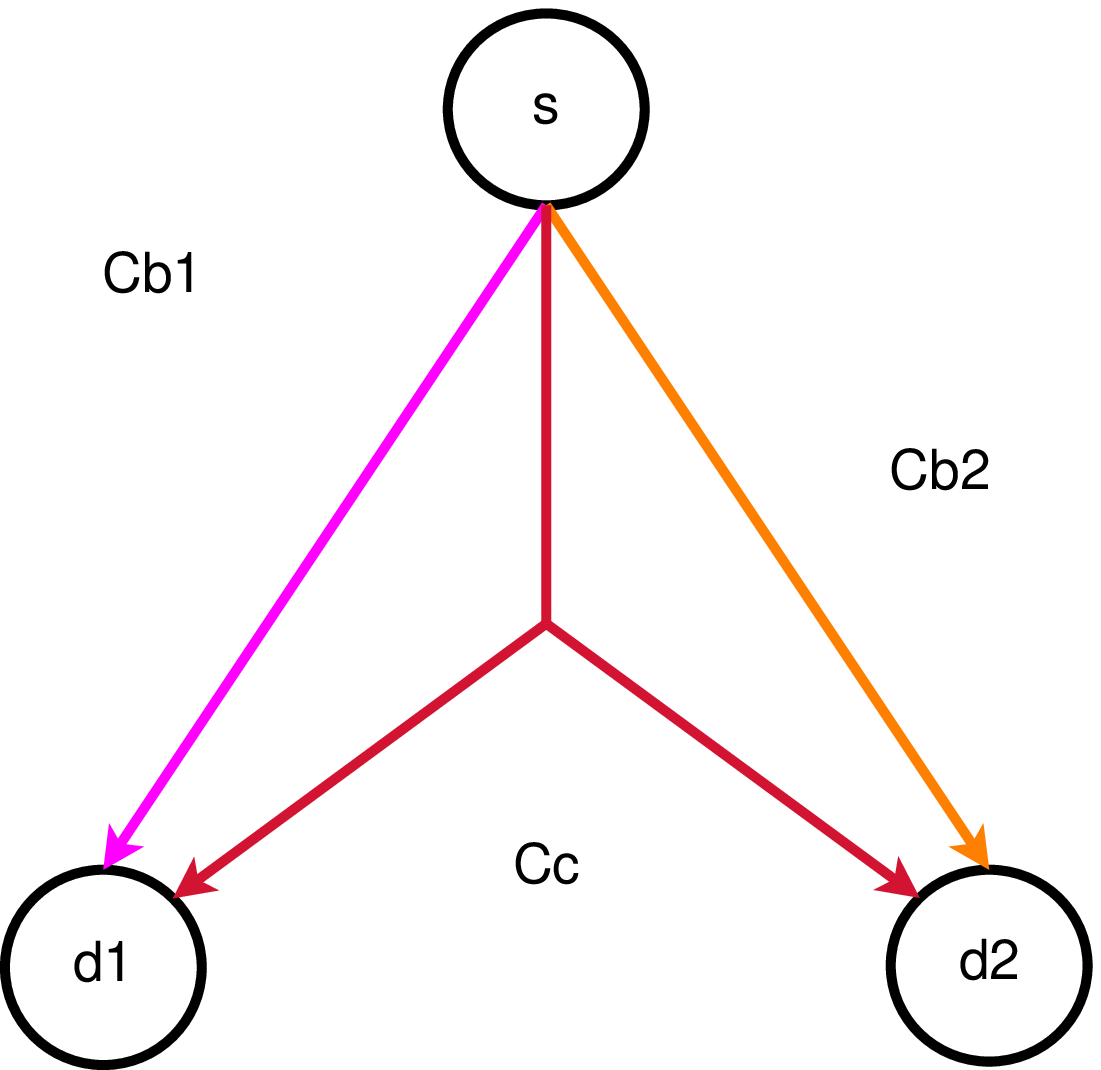}
\label{fig:BC-hypergraph} } \quad \psfrag{s1}[cc][cc]{{\small
$s_1$}} \psfrag{s2}[cc][cc]{{\small $s_2$}}
\psfrag{d}[cc][cc]{{\small $t$}} \psfrag{P1}[cc][cc]{{\small $P_1$}}
\psfrag{P2}[cc][cc]{{\small $P_2$}} \psfrag{N0}[cc][cc]{{\small
$N_0$}} \subfigure[MAC]{
\includegraphics[width=0.25\columnwidth]{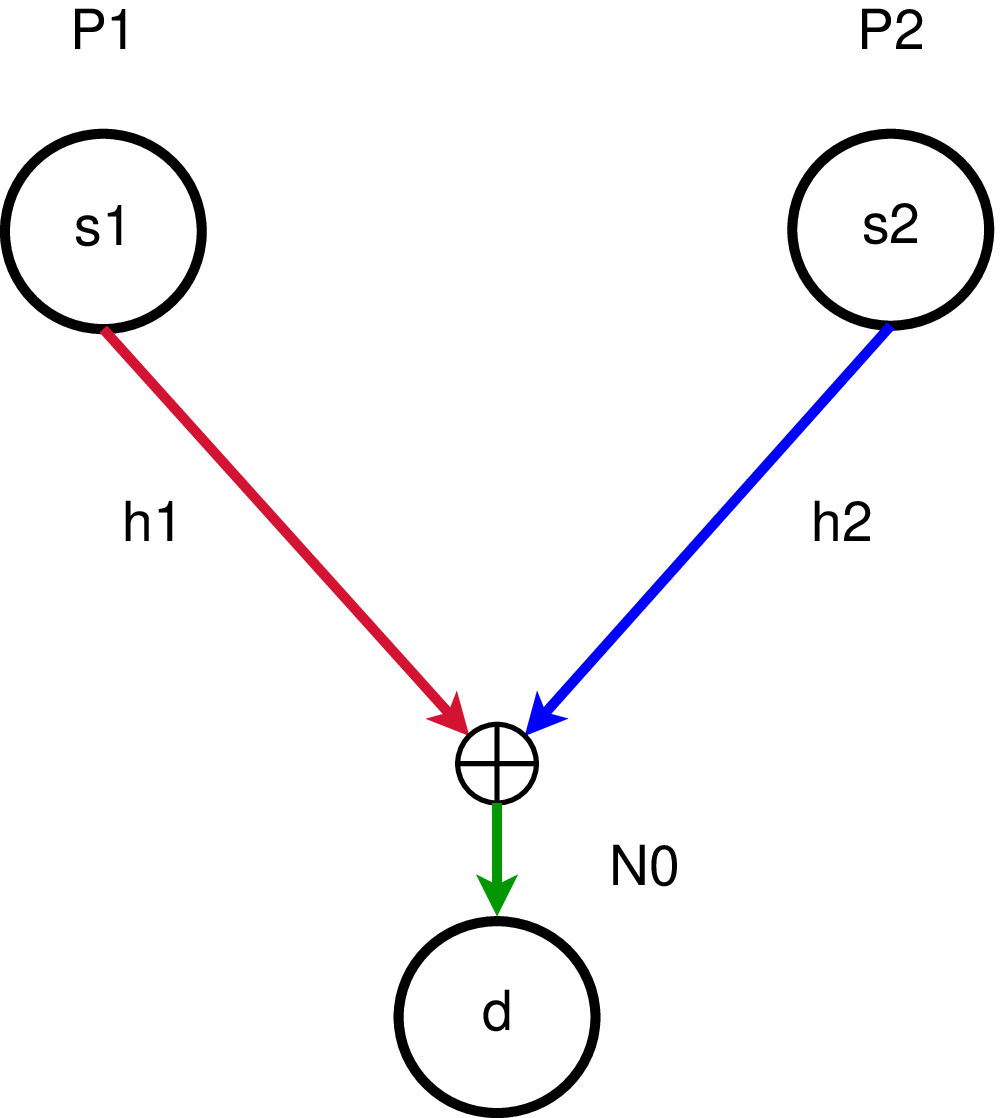}
\label{fig:MAC} } \quad \psfrag{R1}[tc][tc]{{\small $R_1$}}
\psfrag{R2}[br][br]{{\small $R_2$}} \psfrag{C1}[tc][tc]{{\small
\red{$h_1^2\frac{P_1}{N_0}$}}} \psfrag{C2}[cr][cr]{{\small
$C2=h_2^2\frac{P_2}{N_0}$}} \psfrag{C2bis}[tc][tc]{{\small
\blue{$h_2^2\frac{P_2}{N_0}$}}} \subfigure[Wideband MAC equivalent
hypergraph]{
\includegraphics[width=0.26\columnwidth]{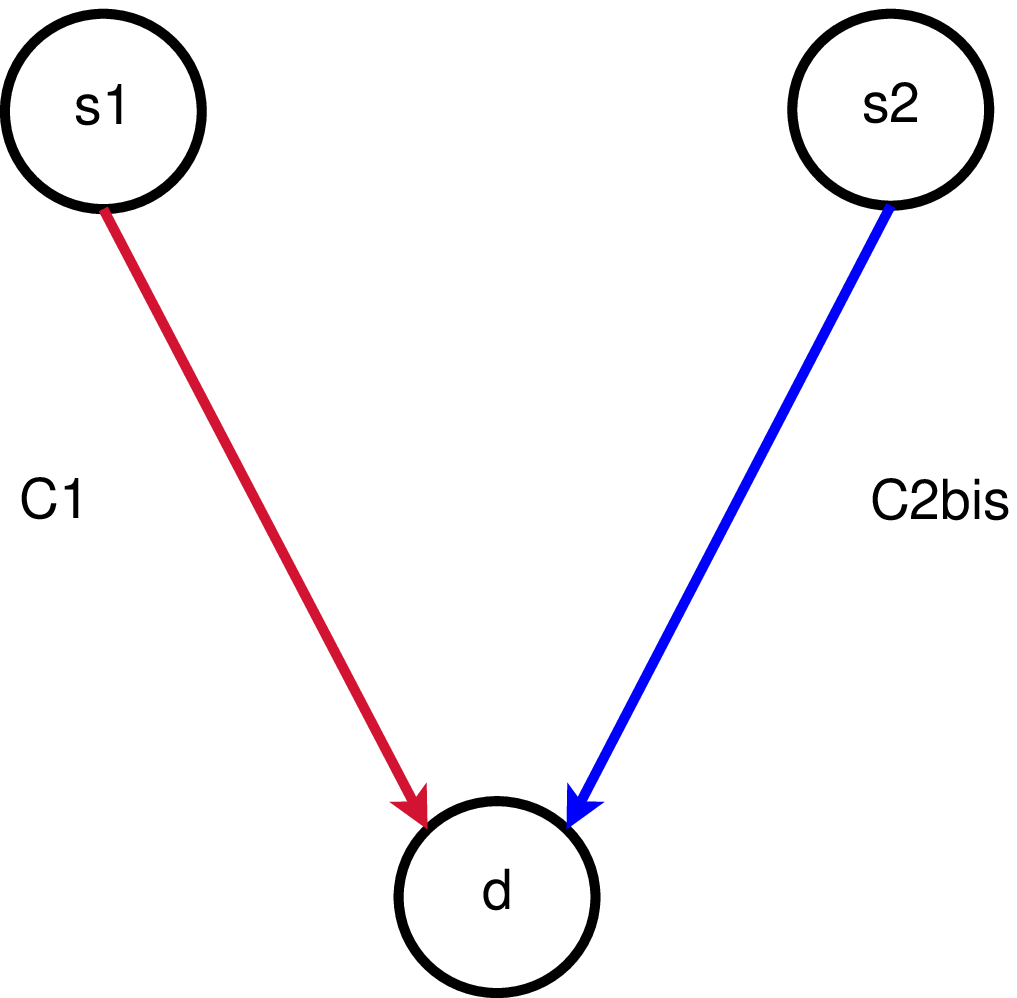}
\label{fig:MAC-hypergraph} } \quad \psfrag{s}[cc][cc]{{\small $s$}}
\psfrag{r}[cc][cc]{{\small $r$}} \psfrag{d1}[cc][cc]{{\small $t_1$}}
\psfrag{d2}[cc][cc]{{\small $t_2$}} \psfrag{Ps}[cc][cc]{{\small
$P_s$}} \psfrag{Pr}[cc][cc]{{\small $P_r$}}
\psfrag{N0}[cc][cc]{{\small $N_0$}} \psfrag{hs1}[cc][cc]{{\small
$h_{1s}$}} \psfrag{hs2}[cc][cc]{{\small $h_{2s}$}}
\psfrag{hsr}[cc][cc]{{\small $h_{rs}$}} \psfrag{hr1}[cc][cc]{{\small
$h_{1r}$}} \psfrag{hr2}[cc][cc]{{\small $h_{2r}$}} \subfigure[BRC]{
\includegraphics[width=0.29\columnwidth]{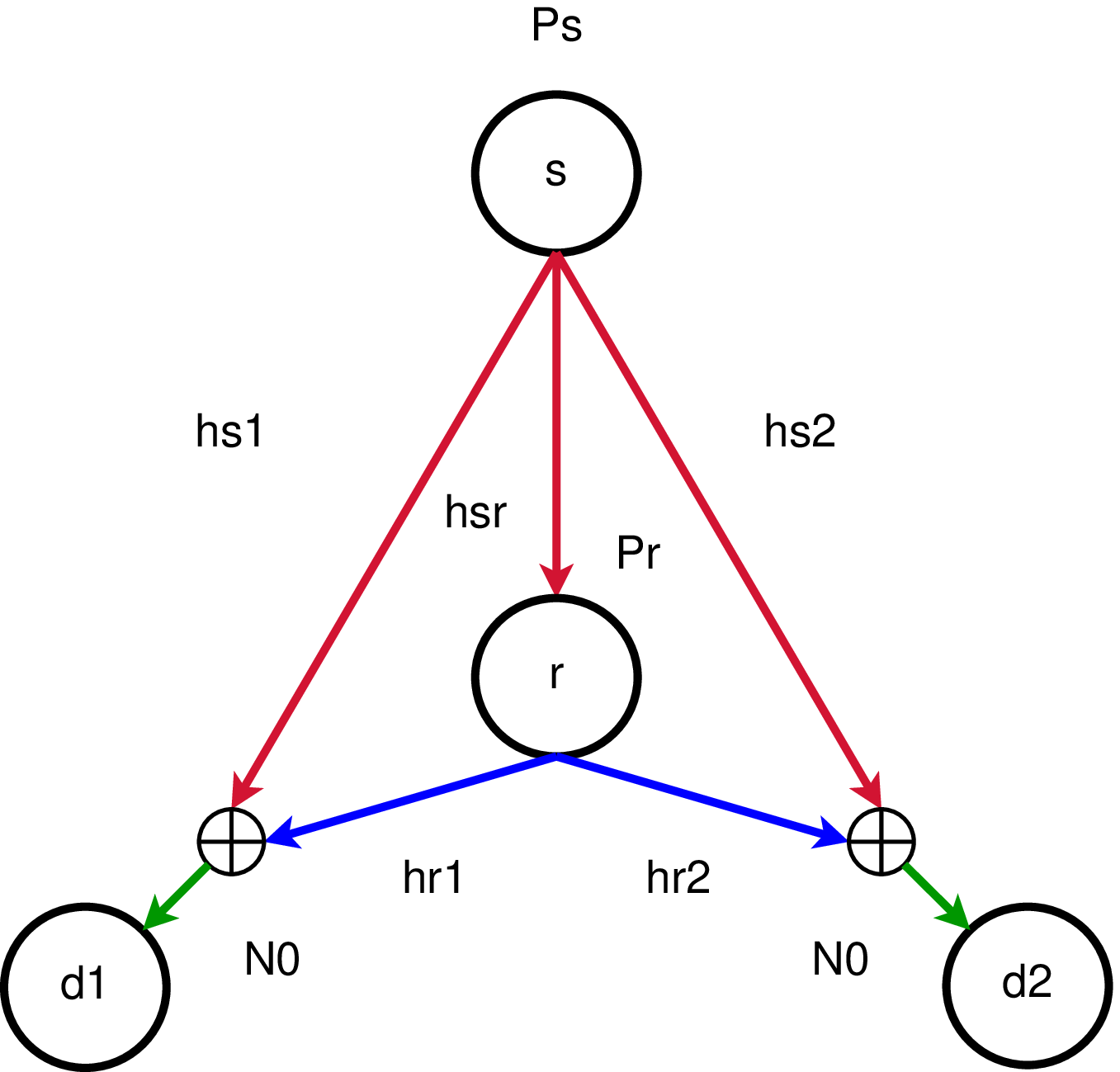}
\label{fig:BRC} } \quad \psfrag{R1}[cc][cc]{{\small
\magenta{$r_0$}}} \psfrag{R2}[cc][cc]{{\small \red{$r_1$}}}
\psfrag{R3}[cc][cc]{{\small \orange{$r_2$}}}
\psfrag{R4}[cc][cc]{{\small \blue{$r_3$}}}
\psfrag{R5}[cc][cc]{{\small \lblue{$r_4$}}} \subfigure[Wideband BRC
achievable hypergraph]{
\includegraphics[width=0.28\columnwidth]{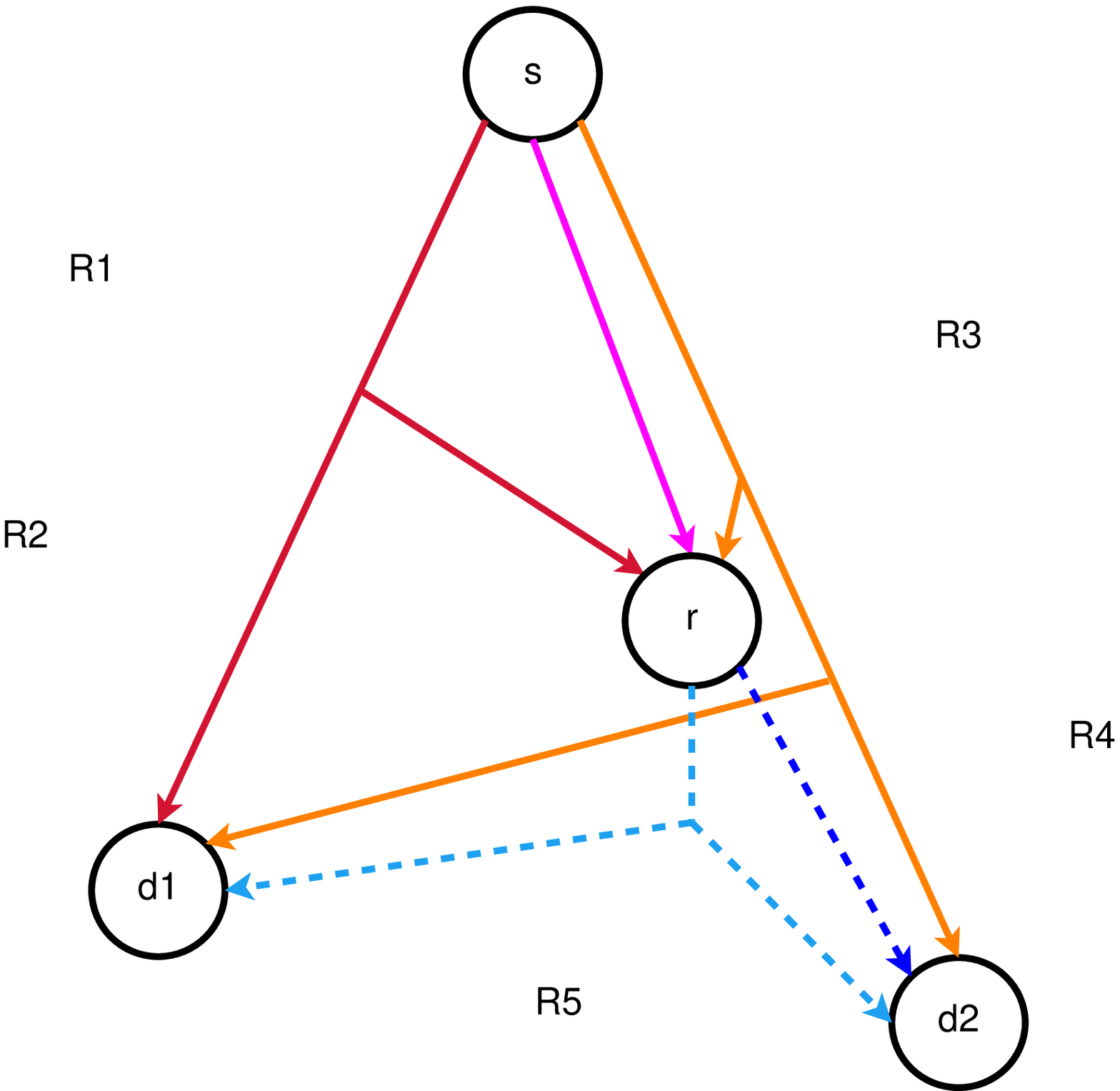}
\label{fig:BRC-hypergraph} } \label{fig:MUrateRegion}
\caption[Wideband MU Channels]{Wideband Multiple User Channels. The
BC rates: $R_1=(1-\beta) h_1^2 \frac{P}{N_0}
\ind_{]h_2^2,+\infty[}(h_1^2)$, $R_2=(1-\beta)  \frac{P}{N_0}
\ind_{[0,h_2^2[}(h_1^2)$, $R_c=\beta \min\{h_1^2,h_2^2\}
\frac{P}{N_0}$. The BRC rates: $r_0=\frac{\beta_0 P_s}{D_{sr}^2
N_0}$, $r_1= \frac{\beta_1 P_s}{D_{st_1}^2 N_0}$, $r_2=\frac{\beta_2
P_s}{D_{st_2}^2 N_0}$,$ r_3=\frac{\mu_1 P_r}{D_{rt_1}^2
N_0}$,$r_4=\frac{\mu_2 P_r}{D_{rt_2}^2 N_0}$. Here, $h$ gives the
path loss and $D_{ij}$ the distance from $i$ to $j$.}
\end{center}
\vspace{-8mm}
\end{figure*}

\vspace{-1mm}
\section{LOW-SNR SYSTEM AND HYPERGRAPH MODEL}\label{sec:SysModel}
\vspace{-1mm}

\subsection{System model and notations}

The network topology is given by a hypergraph $\mathcal{G(N,A)}$,
where $\mathcal{N} = \{s,r,T\}$, and all nodes except $r$ are fixed
on the $2$-D Euclidean plane. $T=\{t_{1},..,t_{n}\}$ denotes the set
of $n=|T|$ receivers ordered in increasing distance from $s$.
$\mathcal{C}$ represents the convex hull of $\{s,T\}$. The multicast
rate from $s$ to $T$ is defined as $R_{sT} \triangleq
\displaystyle\min_{t \in T} (R_{st})$, where $R_{st}$ is the total
rate from $s$ to receiver $t \in T$. $P_{s}$ and $P_{r}=\gamma P_s$
are the total transmit powers of $s$ and $r$, respectively, and
$\gamma >0$ is their ratio.  $D_{uv}$ denotes the Euclidean distance
between nodes $u$ and $v$,  and $\alpha \geq 2$ the path loss
exponent. For a subset $Q \subseteq \mathcal{N} \backslash r$,
define $L_{Q}(\mathcal{C})$ as the point in $\mathcal{C}$, that
minimizes the maximum over the distances between itself and each
node in $Q$, i.e. \vspace{-2mm}
\begin{equation*}
L_{Q}(\mathcal{C}) \triangleq \arg \displaystyle\min_{r \in
\mathcal{C}} \left( \displaystyle\max_{j \in \{Q\}}(D_{rj}) \right).
\hspace{15mm} \mbox{(A)}
\end{equation*}
The value of objective function of the output of Program (A) is
denoted as $D_{Q}$.
\subsection{Low-SNR BC, MAC and BRC hypergraph models}

In
\cite{Thakur-Medard-Globecom2010,Thakur-Fawaz-Medard-Infocom2011},
it was shown that concatenating the low-SNR BC (superposition
coding) and MAC (FDMA) equivalent hypergraph models results in an
achievable hypergraph model for the low-SNR BRC. The rate region of
this model is included in the capacity region of the low-SNR
broadcast relay channel. In fact, even though superposition coding
and FDMA are independently capacity achieving for the low-SNR AWGN
BC and MAC channels respectively, their combination in general is
not capacity achieving for the low-SNR relay channel, and a fortiori
for the low-SNR BRC \cite{Fawaz-Medard-ISIT2010}.

In this section, we briefly recall the equivalent hypergraph models
for the low-SNR BC and MAC, and the achievable hypergraph model for
the BRC \cite{Thakur-Fawaz-Medard-Infocom2011}. Note that in the
low-SNR regime, BC and MAC are \textit{not} limited by interference.

\subsubsection{Low-SNR BC equivalent hypergraph}
Superposition coding is known to achieve the capacity region of the
AWGN BC. In the low-SNR regime, the rates achieved by superposition
coding boil down to the time-sharing region
\cite{Cover-1972,ElGamal-Cover-1980,McEliece-Swanson-1987}. For a
given topology with $|T| = n$ receivers, the hypergraph will contain
at most $n$ hyperarcs with non-zero capacities
\cite{Thakur-Fawaz-Medard-Infocom2011}. Figures~\ref{fig:BC} and
\ref{fig:BC-hypergraph} illustrate the two-destination case.

\subsubsection{Low-SNR MAC equivalent hypergraph}

In the low-SNR regime, interference becomes negligible with respect
to the noise
\cite{Thakur-Medard-Globecom2010,Thakur-Fawaz-Medard-Infocom2011},
and all sources can achieve their point-to-point capacity to the
common destination, like with frequency division multiple access
(FDMA). In the general wideband MAC with $n$ sources, the hypergraph
model consists of $n$ hyperarcs of size $1$ from each source $s_i$,
$i\in \{1,..,n\}$ to the destination with non-zero capacity.
Figures~\ref{fig:MAC} and \ref{fig:MAC-hypergraph} illustrate the
two-source case.

\subsubsection{Low-SNR BRC achievable hypergraph}

We can obtain an achievable hypergraph model of the low-SNR BRC by
simply concatenating the BC and MAC equivalent hypergraphs, as shown
in Figures~\ref{fig:BRC} and \ref{fig:BRC-hypergraph} for the
two-destination case.  As mentioned before, this achievable
hypergraph model is suboptimal in general for the BRC, but the
ability to scale easily to larger and complex networks is one of its
biggest strength.

\section{GEOMETRIC PROPERTIES OF MULTICAST}\label{sec:Geometric}

In this section, we derive the geometric properties of the optimal
relay position maximizing the multicast rate for the BRC. We first
focus on the single destination case of the BRC: the relay channel,
in Section~\ref{sec:RelayChannel}. Then, these preliminary
observations and properties are extended for the general problem
with an arbitrary number of destinations, in
Section~\ref{sec:MultipleDestinations}.

\subsection{Single destination: low-SNR relay channel}\label{sec:RelayChannel}

\begin{figure}[tp]
\begin{center}
\psfrag{s}{$s$} \psfrag{r}{$r$} \psfrag{a}{$(a)$} \psfrag{b}{$(b)$}
\psfrag{c}{$(c)$} \psfrag{d}{$(d)$} \psfrag{e}{$(e)$}
\psfrag{f}{$(f)$} \psfrag{h}{$(h)$} \psfrag{t}{$t$} \psfrag{H}{$H$}
 \psfrag{t1}{$t_{1}$}
 \psfrag{t2}{$t_{2}$}
\psfrag{h1}{$(s,r)$} \psfrag{h2}{$(r,t_{1}t_{2})$}
\includegraphics[width=1\columnwidth]{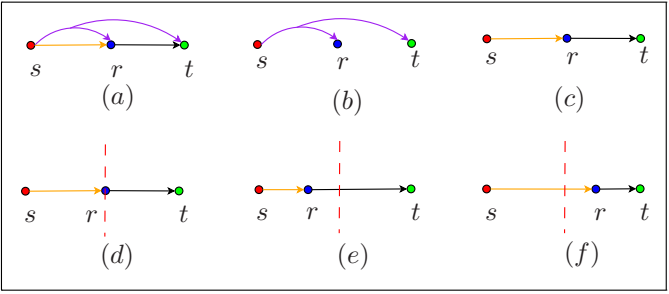}
\end{center}
\vspace{-4mm} \caption{{(a): One receiver case decomposed into two
subgraphs from $s$ to $t$, (b) and (c), respectively. (d): Optimal
position of $r$ for $P_{s}=P_{r}$ and $\alpha =2$, which is at the
perpendicular bisector (red) of line segment $s-t$. (e): Left bias
for $P_{s} < P_{r}$. (f): Right bias for $P_{s}> P_{r}$.}}
\label{fig:1receivercase} \vspace{-6mm}
\end{figure}

Consider the simple network in Figure~\ref{fig:1receivercase}~(a),
with a fixed source $s$, a fixed receiver $t$ and an arbitrarily
positionable relay $r$, where the multicast rate $R_{st}$ from $s$
to $t$ is to be maximized. Naturally, $R_{st}$ depends on the
position of $r$. The achievable hypergraph in
Figure~\ref{fig:1receivercase}~(a) can be broken into two subgraphs,
shown in Figures~\ref{fig:1receivercase}~(b) and (c), which are
essentially the two disjoint paths from $s$ to $t$.

Our claim is that the optimal position of the relay maximizing the
multicast rate from $s$ to $t$ lies on the line segment $s-t$
joining $s$ and $t$, and at this optimal position all the flow
$R_{st}$ is sent through a single path consisting of two hyperarcs,
namely $\{(s,r), (r,t)\}$ shown in
Figure~\ref{fig:1receivercase}~(c). This holds true for any given
pair of power constraints $(P_{s},P_{r}) \succ 0$ and for any path
loss exponent $\alpha \geq 2$. We prove this claim in
Lemmas~\ref{L1} and \ref{L2} hereafter.

We first recall the following lemma from
\cite{Thakur-Fawaz-Medard-Infocom2011}.
\begin{lemma}[Lemma~1 \cite{Thakur-Fawaz-Medard-Infocom2011}]\label{L1}
The optimal position of $r$ maximizing $R_{sT}$ lies inside the
convex hull $\mathcal{C}$.
\end{lemma}

Here, Lemma~\ref{L1} simply implies that the optimal position of $r$
lies on the segment $s-t$.

The rates over the three hyperarcs $\{(s,r),(r,t),(s,rt)\} =
\mathcal{A}$ are given by,
\begin{align}
R_{sr}= \frac{P_{sr}}{D_{sr}^{\alpha}N_{0}}, \hspace{2mm} R_{rt}=
\frac{P_{rt}}{D_{rt}^{\alpha}N_{0}}, \hspace{2mm} R_{srt}= \frac{P_{srt}}{D_{st}^{\alpha}N_{0}}, \label{eq1}\\
P_{sr}+P_{srt} \leq P_{s}, \hspace{2mm} P_{rt} \leq P_{r},
\hspace{20mm} \label{eq2}
\end{align}
where $N_0$ is the noise power spectral density. Note that the
multicast rate is given by $R_{st}=R_{srt}+ \min(R_{sr},R_{rt})$.

\begin{lemma}\label{L2}
The optimal location of $r$ on the segment $s - t$ for a simple BRC
with $\gamma \in (0,\infty)$ and $\alpha \geq 2$ that maximizes the
multicast rate $R_{st}$ satisfies,
\begin{equation}\label{eq:RelayChannelMaxMulticast}
 D_{sr}^{\ast} =\frac{D_{st}}{1+ \sqrt[\alpha]{\gamma}} \hspace{1mm},
\hspace{2mm} D_{rt}^{\ast} =\frac{\sqrt[\alpha]{\gamma}D_{st}}{1+
\sqrt[\alpha]{\gamma}}\hspace{1mm},
\end{equation}
and the optimal (maximized) multicast rate is given by,
\begin{equation}
R^{*}_{st}=\frac{P_{s}}{(D^{*}_{sr})^{\alpha} N_{0}} =\frac{\gamma
P_{s}}{(D^{*}_{rt})^{\alpha} N_{0}}
\end{equation}
where all the flow $R^*_{st}$ is sent over the path
$\{(s,r),(r,t)\}$.
\end{lemma}

In Lemma~\ref{L2} the starred entities refer the optimal values and
for the proof the reader is referred to Appendix A.

Lemma~\ref{L2} essentially gives the position of $r$ in terms of how
far it is from $s$ and $r$ on the segment $s - t$. Also, it provides
the maximized multicast rate $R^{*}_{st}$ that is achieved at this
position. It can be easily seen that the relay position only affects
the rate over the path $\{(s,r),(r,t)\}$. Since the min-cut of the
path $\{(s,r),(r,t)\}$ is strictly larger than the min-cut of the
path $\{(s,rt)\}$, i.e. the rate that can be sent for a unit power
over the former path is strictly larger than the latter path
($R_{srt} < \min(R_{sr},R_{rt})$), the rate over the path
$\{(s,r),(r,t)\}$ should be maximized first by simply maximizing its
min-cut $\min(R_{sr},R_{rt})$ before allocating any power to the
path $\{(s,rt)\}$. The min-cut $\min(R_{sr},R_{rt})$ is maximized at
the position on the segment $s-t$ such that rates over the two
hyperarcs of the path $\{(s,r),(r,t)\}$ become equal, and all the
flow from s to t is transmitted over this path only. The maximized
multicast flow $R^{*}_{st}$ is then simply given by the rates of
either of the two hyperarcs.

Several important conclusions can be drawn from Lemma~\ref{L2}. The
multicast flow optimization can be separated from the determination
of the optimal relay position that maximizes the multicast flow.
Even if the aim is not to maximize the multicast flow (for instance
by simply choosing not to use all the source and relay powers),
Lemma~\ref{L2} still gives the most suitable relay position for any
feasible multicast rate $R_{st} \leq R_{st}^{*}$. At the same time,
the algorithmic style intuitive proof arguments in the previous
paragraph indicate that upon computing the optimal relay position,
the multicast rate maximization problem could be casted as a
straightforward linear program resulting in a simple power
allocation scheme maximizing the multicast rate. This fact will
prove handy for the general case with arbitrary number of
destinations. On the other hand, we observe the dependency of the
optimal relay position on the constants $\alpha$ and $\gamma$. If
$\gamma = 1$ i.e. $P_{s}=P_{r}$, the optimal relay position is
always at the mid-point of the segment $s-t$ for any value of
$\alpha \geq 2$. When $\gamma \neq 1$, there will be a natural bias
on the optimal position of $r$ either towards $s$ or $t$, depending
on the value of $\gamma$. This bias will also depend on the value of
$\alpha$. Figure~\ref{fig:1receivercase}(e) and
\ref{fig:1receivercase}(f) show the bias effect.

\subsection{Multiple destinations}\label{sec:MultipleDestinations}

In this subsection, we extend the simple geometric insights
developed in Section~\ref{sec:RelayChannel} for a single destination
to the general case of an arbitrary number of destinations $|T|=n$.

Let us first note the following. For a given hypergraph
$\mathcal{G(N,A)}$,
 and a fixed position of $r$, we have at most
$(n+1)+(n)$ hyperarcs in the system, i.e. $|\mathcal{A}|=2n+1$. The
former $(n+1)$ are source hyperarcs, emanating from $s$ to the nodes
in $\mathcal{N} \backslash s$ and the latter $n$ are the relay
hyperarcs, emanating from $r$ to all $T$. Also, for any given
position of $r$ there always exist at least two paths that will span
all the receiver set $T$, namely $\{(s,T)\}$ (or $\{s,t_1..t_n\}$)
and $\{(s,T_{1}),(r,T_{2})\}$ (where $r \in T_{1}$ and $T_{1} \cup
T_{2}=\{r,T\}$).

Now, consider that each hyperarc $(i,J) \in \mathcal{A}$ is
associated with a continuous function $f_{iJ}(P_{i}^+,D_{iJ}^-):\R^2
\longrightarrow \R$, that is a monotonically increasing in the
transmit node's power $P_{i}$ and monotonically decreasing in the
distance $D_{iJ}$, where $D_{iJ}$ is the Euclidean distance between
the transmit node $i$ and the farthest receiver node $j \in J$ (from
$i$) spanned by the hyperarc. Then the following theorem holds true.

\begin{theorem}\label{T1}
Given a hypergraph $\mathcal{G(N,A)}$ and the associated rate
functions $f_{iJ}(P_{i}^+,D_{iJ}^-):\R^{2} \longrightarrow \R$ for
each hyperarc in $\mathcal{A}$, at the optimal position maximizing
the multicast rate $R_{sT}$ one of the two multicast flow
characteristics holds:
\begin{enumerate}[(i)]
 \item all the optimal flow $R^{*}_{sT}$ goes through at most two paths $\{(s,T_{1},(r,T_{2})\}$ and $\{(s,T)\}$,
in succession.
\item all the optimal flow $R^{*}_{sT}$ can be arbitrarily split between the two paths  $\{(s,T)\}$ and $\{(s,T_{1}),(r,T_{2})\}$.
\end{enumerate}
\end{theorem}
\vspace{1mm}

For the proof of Theorem~\ref{T1}, refer to Appendix B.

Theorem~\ref{T1} partially generalizes Lemma~\ref{L2}. We say
partially, because on one hand, Theorem~\ref{T1} establishes the
important multicast flow characteristics at the optimal relay
position, but it does not provide a simple numerical result that
determines the optimal relay location (like Lemma~\ref{L2}). Note
that, for a given relay position there could be multiple paths from
$s$, through $r$, to all $T$, but in the Theorem~\ref{T1} by path
$\{(s,T_{1}),(r,T_{2})\}$ we mean the path from $s$, through $r$, to
all $T$ that has the highest min-cut among all the paths from $s$,
through $r$, to all $T$. Intuitively, Theorem~\ref{T1} states that
only those paths will contain the multicast flow from $s$ to the
receiver set $T$ that serve all $T$, namely $\{(s,T)\}$ and
$\{(s,T_{1}),(r,T_{2})\}$. All other path that serve proper subsets
of $T$ will carry no flow as they do not contribute to the multicast
flow and among all the paths serving all $T$ through $r$, only the
path with the highest min-cut will carry the multicast flow.
 This fact is a simple yet fundamental consequence of the definition of multicast.

Theorem~\ref{T1} reveals a lot about the nature of multicast flow
over a hypergraph. The dependence of relay position on the rate of
only a single path $\{(s,T_{1}),(r,T_{2})\}$ reduces the problem to
its core by removing the clutter away. In other words, now we only
need to worry about the maximization of the flow over this single
path and the relay position that maximizes the flow over this path
also maximizes the multicast flow $R_{sT}$. This result of
Theorem~\ref{T1} motivates a pure geometric interpretation of the
problem. If we imagine the two hyperarcs $(s,T_{1})$ and $(r,T_{2})$
to be two circles $C_{s}$ and $C_{r}$ centered at $s$ and $r$ with
radii $\pi_{s}$ and $\pi_{r}$, respectively, then the optimal relay
positioning problem could be stated as: \emph{For a given
$\mathcal{G(N,A)}$, find the point in $\mathcal{C}$ such that when
$r$ is positioned at this point,
$\max(\sqrt[\alpha]{\gamma}\pi_{s},\pi_{r})$ is minimized while $r
\in C_{s}$ and the region of union of two circles $C_{\cup}=C_{s}
\cup C_{r}$ encompasses all $T$}.

At first, it seems plausible to try a simple (preferably convex)
optimization framework to compute such a point, but the condition
that the two circles must encompass all $\mathcal{N}$ brings in
discreteness, which we avoid for obvious reasons. In contrast, we
propose a simple (polynomial time) algorithm to compute such point
in the next sections. Once the optimal relay position is obtained,
obtaining optimal power allocations (for $s$ and $r$) maximizing the
multicast rate boils down to solving a simple linear program
involving only two paths. We divide the development of this
algorithm into two cases of $\gamma =1$ and $\gamma \in (0,\infty)$.
The case of $\gamma=1$ is easy to understand and holds importance in
its own right. In addition it develops the basic intuition for the
proposed algorithm and leaves the extension to the case of all
values of $\gamma \in (0,\infty)$, as straightforward.

\vspace{-1mm}
\section{$(P_{s}=P_{r})$ - CASE AND ALGORITHM}\label{sec:eq}
\vspace{-1mm} In this section, we have $\gamma =1$ and $\alpha \geq
2$ for a given $\mathcal{G(N,A)}$  on the $2$-D Euclidean plane. The
optimal relay positioning problem stated geometrically in the
previous section simply boils down to finding the point in
$\mathcal{C}$ such that $\max(\pi_{s},\pi_{r})$ is minimized while
$r \in C_{s}$ and $C_{\cup}$ encompasses all $T$. We divide the
problem in the following two cases based on the topology of the
given $\mathcal{G(N,A)}$.

\vspace{-1mm}
\subsection{$s-t_{n}$ mid-point case}
\vspace{-1mm}
\begin{lemma}\label{L3}
If $r$ is placed at the mid-point of $s-t_n$ such that the hyperarcs
$C_{s}$ and $C_{r}$ each with radii $\frac{D_{st_n}}{2}$ span all
$T$, then it is the optimal relay position maximizing $R_{sT}$.
\end{lemma}\vspace{1mm}

The proof of Lemma~\ref{L3} is a straightforward generalization of
Lemma~\ref{L2} and therefore is omitted. Intuitively, Lemma~\ref{L3}
simply states that since the farthest node (from $s$) $t_{n}$ is
also the limiting node for maximizing $R_{sT}$, if the rate is
maximized only to $t_{n}$ while guaranteeing it to all other nodes
in $T$, then this maximizes $R_{sT}$ as well. This means that if $r$
is placed at the mid-point of the segment $s-t_n$ (as this position
maximizes the rate to $t_n$ only) and if the two hyperarcs of the
path $\{(s,r),(r,t_n)\}$ ($\{C_{s},C_{r}\}$) span all $T$, then
clearly this is the relay position that maximizes $R_{sT}$.

\vspace{-1mm}
\subsection{General Case}
\vspace{-1mm} In this case we tackle all topologies and case $A$
becomes a special case of it. Recall that, the entity
$L_{Q}(\mathcal{C})$ represents the coordinates of the point which
is the argument of the objective function of the output of program
(A), and $D_{Q}$ is the value of the objective function of the
output of program (A).

\vspace{1mm}
\underline{\textbf{\emph{Optimal relay positioning Algorithm (ORP)}}} \\
Given: $\mathcal{G(N,A)}$.
\begin{enumerate}
\item Compute $l_{0}=L_{\{\mathcal{N} \backslash r\}}(\mathcal{C})$ and build the set
$\mathbf{N_0}=\{t \in T|D_{st} < D_{l_{0}t}  \with D_{l_{0}t} >
D_{sl_{0}}\}=\{t'_1,..,t'_{m}\}$ in increasing order of distance
from $s$. If $\mathbf{N_0}=\{\emptyset\}$, declare $l_{0}$ as the
optimal relay position and quit, else go to step $2$.
\item Build the set $\mathbf{N_{1}}=\{\mathcal{N} \backslash (r,\mathbf{N_0})\}$ and compute the point $l_{1}=L_{\mathbf{N_{1}}}(\mathcal{C})$.
Form the hyperarcs $C_{s}$ and $C_{l_1}$ of radii $D_{sl_1}$ and
$D_{\mathbf{N_{1}}}$, respectively. If $C_{\cup}=C_{s} \cup
C_{l_{1}}$ encompasses all $T$, output $l_{1}$ as the optimal relay
position and quit, else go to step $3$.
\item Reform the hyperarc $C_{s}$ of radius $D_{st'_m}$ and build the set $\mathbf{N_{2}}=\{t \in T| D_{st} > D_{st'_m}\}$ and compute $l_{2}=L_{\mathbf{N_{2}}}(\mathcal{C})$. Declare $l_2$ as the optimal relay position and quit.
\end{enumerate}

Algorithm ORP is a straightforward set of basic and intuitive
computational steps based on the properties of the point
$l_{0}=L_{\mathcal{N} \backslash r}(\mathcal{C})$. If there exist no
node $t' \in T$ such that $t' \notin C_{s}$ and $D_{st'} <
D_{l_{0}t'}$
 (i.e. set $\mathbf{N_0}$
is empty), that can be directly reached by $s$ rather than by a path
through $r$, then $l_{0}$ is certainly the optimal relay position.
In contrast, if the set $\mathbf{N_0}$ is not empty, then there
exist at least one receiver node in the system that influences the
computation of the optimal relay position but can be served directly
by $C_s$. Therefore, either the nodes in $\mathbf{N_0}$ can be
removed from the computation of the optimal relay position ($l_{1}$
in Step $2$) and $\max(\pi_{s},\pi_{r})$ can be further reduced or
we could reform the hyperarc $C_{s}$ with radius $D_{st'_m}$ (where,
$t'_m$ is the farthest node in $\mathbf{N_0}$ from $s$) and then
computing the point $l_2$ for the nodes that were not covered by
$C_s$ and thus reducing the value of $\max(\pi_{s},\pi_{r})$. Note
that, Algorithm ORP categorizes all possible topologies of the given
$\mathcal{G(N,A)}$ in three steps and there is no underlying
iterative process. This makes algorithm ORP behave like a numerical
formula, which we originally wanted from Theorem~\ref{T1}.

We leave the formal proof that $ORP$ always outputs the optimal
relay position maximizing $R_{sT}$ to Appendix C and extend this
simple approach in a straightforward manner to the case of all
values of $\gamma \in (0,\infty)$ in the next section.

\section{$P_{s} \neq P_{r}$- CASE AND ALGORITHM}\label{sec:neq}

In this section, we consider $\gamma \in (0,\infty)$ for a given
$\mathcal{G(N,A)}$ and $\alpha \geq 2$. Almost all the theory
developed in Section~\ref{sec:eq} simply transcends to this section,
with certain notable differences. Mainly, that when $\gamma \neq 1$
it gives rise to a bias in the positioning of $r$ ( ref.
Figure~\ref{fig:1receivercase}(e) and \ref{fig:1receivercase}(f)).
Taking into account the bias while computing the optimal relay
position will be the main enhancement in this section. Likewise
previously, we first consider the $s-t_n$ case. \vspace{-1mm}
\subsection{$s-t_{n}$ case}
\vspace{-1mm}
\begin{lemma}\label{L4}
Given $\mathcal{G(N,A)}$, if $r$ is placed on $s-t_n$ at a distance
of $D_{sr}=\frac{D_{st_n}}{1+\sqrt[\alpha]{\gamma}}$ from $s$, such
that $r \in C_{s}$ and $C_{\cup}=C_{s} \cup C_{r}$  spans all $T$,
then it is optimal relay position that maximizes $R_{sT}$.
\end{lemma}

The line of argument for the proof of Lemma~\ref{L3} (using
Lemma~\ref{L2}) could be simply generalized for Lemma~\ref{L4}.


\subsection{General Case}
In this case, like in Section~\ref{sec:eq}, we generalize to all
topologies. As we know, that the values of $\gamma$ (when not equal
to $1$) and $\alpha$ inflict the bias on the relay position. The
main difference in case of $P_{s} \neq P_{r}$ is the computation of
the point $l_{i}=L_{Q}(\mathcal{C})$ ($i=\{0,1\}$), given by,
\begin{equation*}
l_{i}=L_{Q}(\mathcal{C}) \triangleq \arg\displaystyle\min_{i \in
\mathcal{C}} \left( \displaystyle\max_{(j \in Q \backslash s)}
\left(\sqrt[\alpha]{\gamma} D_{si},D_{ij}\right) \right).
\hspace{5mm} \mbox{(B)}
\end{equation*}
and the computation of the set $\mathbf{N_0}=\{t \in T|
\sqrt[\alpha]{\gamma} D_{sl_0}
> D_{l_0t}\}=\{t'_1,..,t'_m\}$, in the Algorithm ORP. Program (B) and the set $\mathbf{N_0}$ takes into account
the bias induced by the differences in the transmit power of the
source and relay and the value of $\alpha$. The rest of the
algorithm remains the same.

\begin{figure}[tp]
\begin{center}
\psfrag{s}{$s$} \psfrag{r}{$r$} \psfrag{a}{$(a)$} \psfrag{b}{$(b)$}
\psfrag{c}{$(c)$} \psfrag{tn}{$t_n$} \psfrag{t3}{$t_3$}
\psfrag{C}{$C_{\cap}$}
\includegraphics[width=1.0\columnwidth]{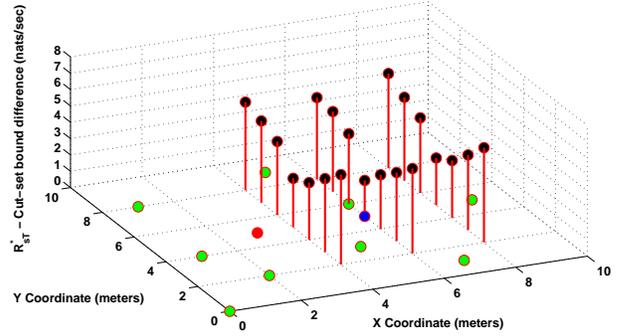}
\end{center}
\vspace{-4mm} \caption{{$|T|=9$ case with green receivers, red
source and blue as the optimal relay position. The optimal $R_{sT}$
and cut set bound difference (in nats/sec) is calculated for $21$
positions and is the lowest at the optimal relay position (blue). We
assume $\frac{P_{s}}{N_{0}}=\frac{P_{r}}{N_{0}}=1$ (normalized) and
$\alpha=4$. }} \label{fig:Fig3} \vspace{-6mm}
\end{figure}

Now that we have an efficient algorithm for computing the optimal
relay position, we can be more ambitious to assess the standing of
our work in a more theoretical sense. One of the important
consequences of this work that signifies its theoretical importance
is shown in Figure~\ref{fig:Fig3}. We computed the difference
between the optimal multicast rate $R^{*}_{sT}$ (for a given
position of $r$) and the cut set bound for $|T|=9$ receiver nodes
network at $21$ interesting positions, including the optimal relay
position computed by the Algorithm ORP. At the optimal relay
position (blue point), this difference is minimized, confirming the
fact that the optimal relay position not only results in gains but
the maximized multicast rate is theoretically closest to the cut-set
bound at the optimal relay position in our framework.

It is worth mentioning that the theory developed in this paper well
transcends to the low-SNR fading channels , which we do not discuss
here but can be easily generalized from the results of
\cite{Thakur-Medard-Globecom2010} and \cite{Fawaz-Medard-ISIT2010}.

%

\section{CONCLUSION} \label{sec:Conclusion}

We list the important deductions from our work in the following
points.
\begin{enumerate}
 \item The problem of optimal relay positioning to maximize the multicast rate for the
 achievable hypergraph model of low-SNR networks using superposition
coding and FDMA, can be decoupled from flow optimization and casted
as a simpler geometric problem, as opposed to a complex network
optimization approach of \cite{Thakur-Fawaz-Medard-Infocom2011}.
\item The geometric properties of multicast are innately simple
and provide interesting insights for relay positioning problem. This
is largely due to the fact that all the multicast flow is pushed
over at most two paths which is a direct consequence of the
definition of the multicast flow, and  this results in simple
geometric interpretation.
\item Importantly, the benefits of determining the optimal relay
position are substantiated by the fact that the difference between
the maximized multicast rate and the cut-set bound at the optimal
position is minimized.
\end{enumerate}

We now outline, what we think are certain important future
directions our work could take. The geometric properties of
multicast give great insights and are surprisingly easy to work
with. This motivates us to ask further, whether is it possible to
apply the simple techniques of our work for the optimal relay
positioning problem to moderate and high-SNR regimes that are
interference limited. Another natural and interesting dimension is
to look at the possibility of extending this work to multicommodity
flows.

\vspace{-1.5mm}

\bibliographystyle{./../../Biblio/IEEEtran}

\bibliography{./../../Biblio/IEEEabrv,./../../Biblio/bibLowSNR,./../../Biblio/bibNC,./../../Biblio/bibCellularStd,./../../Biblio/bibOptim}

\begin{thebibliography}{1}
\providecommand{\url}[1]{#1}
\csname url@rmstyle\endcsname
\providecommand{\newblock}{\relax}
\providecommand{\bibinfo}[2]{#2}
\providecommand\BIBentrySTDinterwordspacing{\spaceskip=0pt\relax}
\providecommand\BIBentryALTinterwordstretchfactor{4}
\providecommand\BIBentryALTinterwordspacing{\spaceskip=\fontdimen2\font plus
\BIBentryALTinterwordstretchfactor\fontdimen3\font minus
  \fontdimen4\font\relax}
\providecommand\BIBforeignlanguage[2]{{%
\expandafter\ifx\csname l@#1\endcsname\relax
\typeout{** WARNING: IEEEtran.bst: No hyphenation pattern has been}%
\typeout{** loaded for the language `#1'. Using the pattern for}%
\typeout{** the default language instead.}%
\else
\language=\csname l@#1\endcsname
\fi
#2}}

\bibitem{Thakur-Fawaz-Medard-Infocom2011}
M.~Thakur, N.~Fawaz, and M.~M{\'e}dard, ``Optimal relay location and power
  allocation for low snr broadcast relay channels,'' in \emph{Proc. IEEE
  International Conference on Computer Communications, {INFOCOM} 2011,
  Shanghai, China}, Apr. 2011.

\bibitem{Thakur-Medard-Globecom2010}
M.~Thakur and M.~M{\'e}dard, ``On optimizing low {SNR} wireless networks using
  network coding,'' in \emph{Proc. IEEE Global Communications Conference,
  {Globecom} 2010, Miami, FL, USA}, Dec. 2010.

\bibitem{Fawaz-Medard-ISIT2010}
\BIBentryALTinterwordspacing
N.~Fawaz and M.~M{\'e}dard, ``On the non-coherent wideband multipath fading
  relay channel,'' in \emph{Proc. IEEE International Symposium on Information
  Theory, {ISIT} 2010, Austin, TX, USA}, June 2010. [Online]. Available:
  \url{http://arxiv.org/abs/1002.3047}
\BIBentrySTDinterwordspacing

\bibitem{Cover-1972}
T.~M. Cover, ``Broadcast channels,'' \emph{{IEEE} Trans. Inform. Theory},
  vol.~18, no.~1, Jan. 1972.

\bibitem{ElGamal-Cover-1980}
A.~E. Gamal and T.~M. Cover, ``Multiple user information theory,''
  \emph{Proceedings of the IEEE}, vol.~68, no.~12, pp. 1466--1483, Dec. 1980.

\bibitem{McEliece-Swanson-1987}
R.~McEliece and L.~Swanson, ``A note on the wide-band gaussian broadcast
  channel,'' \emph{{IEEE} Trans. Commun.}, vol.~35, no.~4, pp. 452--453, Apr.
  1987.

\bibitem{Thakur-Fawaz-Medard-arXivISIT2011}
\BIBentryALTinterwordspacing
M.~Thakur, N.~Fawaz, and M.~M{\'e}dard. (2011) On the geometry of wireless
  network multicast in 2-{D}. [Online]. Available: \url{http://arxiv.org/}
\BIBentrySTDinterwordspacing

\end{thebibliography}

\newpage

\appendices
\section{Proof of \emph{Lemma~\ref{L2}}}\label{ap:Proof-L2}

\begin{IEEEproof}
We only consider the positions in the interior of the segment
$s-t$. Then, the multicast rate is given by
\begin{equation}\label{L2e1}
\begin{split}
& R_{st}
=R_{srt}+ \min(R_{sr},R_{rt})\\
&= \frac{\lambda P_{s}}{D_{st}^{\alpha/2}N_{0}} +\min
\left(\frac{(1-\lambda) P_{s}}{D_{sr}^{\alpha}N_{0}}, \frac{
{\gamma}P_{s}}{D_{rt}^{\alpha}N_{0}} \right)\\
&= \frac{P_s}{N_0}\min \left(  \lambda
\left(\frac{1}{D_{st}^{\alpha}}-\frac{1}{D_{sr}^{\alpha}}\right)
+\frac{1}{D_{sr}^{\alpha}},\lambda \frac{1}{D_{st}^{\alpha}} +
\frac{ {\gamma}}{D_{rt}^{\alpha}} \right).
\end{split}
\end{equation}
By assumption, we have $D_{st} > \max(D_{sr},D_{rt})$. Thus, in the
minimization of~(\ref{L2e1}), the first and the second term are
respectively decreasing and increasing affine functions of
$\lambda$. Two cases can occur. If $\sqrt[\alpha]{\gamma} D_{sr}
\geq D_{rt}$, then the second term is always larger than the first
term, which consequently is the minimum of the two. The first term
decreases in $\lambda$, thus $R_{st}$ is maximized for $\lambda=0$.
Else, if $\sqrt[\alpha]{\gamma} D_{sr} \leq D_{rt}$, the two affine
functions intersect in the interval $[0,1]$ at $\lambda = 1 -
\frac{\gamma D_{sr}^{\alpha}}{D_{rt}^{\alpha}}$. The multicast rate
$R_{st}$ is maximized at this intersection. Note that for the
position of $r$ satisfying $\sqrt[\alpha]{\gamma} D_{sr} = D_{rt}$,
both solutions match: $\lambda = 1 - \frac{\gamma
D_{sr}^{\alpha}}{D_{rt}^{\alpha}} =0$.

By Lemma~\ref{L1}, the relay position maximizing the multicast rate lies on segment $s-t$. Then, we can write
\begin{equation}\label{eq:onSegment}
D_{st}=D_{sr}+D_{rt},
\end{equation}
and the relay position is simply determined by the distance
$D_{sr}$. Using (\ref{eq:onSegment}), the conditions
$\sqrt[\alpha/2]{\gamma} D_{sr} \lessgtr D_{rt}$ in Lemma~\ref{L2}
can be rewritten in function of $D_{sr}$ as
\begin{equation}\label{eq:DsrCases}
D_{sr} \lessgtr \frac{D_{st}}{1+ \sqrt[\alpha]{\gamma}}.
\end{equation}
Given the optimal power allocation $\lambda^\ast$, and using
(\ref{eq:DsrCases}), the multicast rate $R_{st}$ can be rewritten as
the following function of $D_{sr}$
\begin{equation}\label{eq:RelayChannelMulticast}
\begin{split}
 R_{st}=
 \left\{
 \begin{array}{ll}
  \frac{P_s}{D_{st}^{\alpha} N_0} \left( 1 + \gamma \frac{D_{st}^{\alpha} -D_{sr}^{\alpha}}{(D_{st}-D_{sr})^{\alpha}} \right),
& \mbox{if }D_{sr} \leq \frac{D_{st}}{1+ \sqrt[\alpha]{\gamma}}; \\
   \frac{ P_{s}}{D_{sr}^{\alpha}N_{0}}, & \mbox{if }  D_{sr} \geq \frac{D_{st}}{1+ \sqrt[\alpha]{\gamma}}.
 \end{array}
 \right.
\end{split}
\end{equation}
From (\ref{eq:RelayChannelMulticast}), it can be seen that $R_{st}$
is an increasing function of $D_{sr}$ over $(0,\frac{D_{st}}{1+
\sqrt[\alpha]{\gamma}}]$, and then a decreasing function of $D_{sr}$
over $[\frac{D_{st}}{1+ \sqrt[\alpha]{\gamma}},D_{st})$. Therefore,
the multicast rate is maximized at the border between these two
intervals: $D_{sr}^\ast=\frac{D_{st}}{1+ \sqrt[\alpha]{\gamma}}$.
Substituting $D_{sr}^\ast$ in (\ref{eq:RelayChannelMulticast})
yields $R_{st}^\ast$. This completes the proof of the Lemma.
\end{IEEEproof}

\section{Proof of \emph{Theorem~\ref{T1}}}\label{ap:Proof-T1}
\vspace{1mm}
A simple assimilation of the basic graph theoretic and Euclidean geometric concepts helps form the fundamental reasoning for the proof.
Let's assume that the hypergraph $\mathcal{G(N,A)}$ is given with the constant $\gamma \in (0,\infty)$ (where, $P_{r}=\gamma P_{s}$)
and each hyperarc $(i,J) \in \mathcal{A}$ is associated with any continuous rate function
$f_{iJ}(P^{+}_{i},D^{-}_{iJ}):\R^2 \longrightarrow \R$, that is monotonically increasing in the transmit power of the emanating node $i$ of the
hyperarc and is monotonically decreasing in the distance $D_{iJ}$ (between the transmit node $i$ and the farthest node $j \in J$ from $i$). We notice that
there are only two transmitters in the system $s$ and $r$ and the multicast rate from $s$ to the receiver set $T$ is defined as
$R_{sT}=\displaystyle\min_{(t \in T)} (R_{st})$, where $R_{st}$ is the total rate received by the receiver $t \in T$.

\vspace{2mm}
\begin{IEEEproof}
To prove the theorem, we first notice that for a given position of $r$ there are at least two paths, namely certain paths of the type
$\{(s,T_{1}),(r,T_{2})\}$  (where, $T_{1} \cup T_{2}=T$) and the path $\{(s,T)\}$, that span the whole receiver
set $T$. Any other path, that only serves the proper subsets of $T$ does not count in contribution to the multicast rate $R_{sT}$.

Among all the paths from $s$ to $T$, that go through $r$ (i.e. of the type $\{(s,T_{1}),(r,T_{2})\}$, where, $T_{1} \cup T_{2}=T$),
only the path with highest min-cut
contributes to the multicast flow $R_{sT}$. Let us denote this path as $\{(s,T'_{1}),(r,T'_{2})\}$, where $T'_{1} \cup T'_{2}=T$.
Once the min-cut of the path $\{(s,T'_{1}),(r,T'_{2})\}$ is reached, considering it has the highest min-cut among the paths that span all
$T$ through $r$, no flow can be sent over any other path of the type $\{(s,T_{1}),(r,T_{2})\}$. This is true because when the min-cut of
the path $\{(s,T'_{1}),(r,T'_{2})\}$ is achieved (for a fixed position of $r$) either $P_{s}$ is consumed or $P_{r}$ is consumed.
If $P_{s}$ is consumed, no more multicast flow can be pushed, and if $P_{r}$ is consumed before $P_{s}$ then rest of the flow have to be pushed over
the path $\{(s,T)\}$ (not involving $r$). On the other hand for a given position of $r$,
if the min-cut of the path $\{(s,T'_{1}),(r,T'_{2})\}$ is strictly less than of
$\{(s,T)\}$, then all the multicast flow must be sent over the path $\{(s,T)\}$.

This implies that for any given position of $r$, all the multicast flow must be sent over at most these two paths. Now,
we can write down the min-cut of the multicast flow as
\begin{equation*}
\begin{split}
 R_{sT}=&f_{sT}(P^{+}_{s},D_{sT}^{-}) + \\
&\min(f_{sT'_{1}}(P^{+}_{s},D_{sT'_{1}}^{-}),f_{rT'_{2}}(P^{+}_{r},D_{rT'_{2}}^{-})).
\end{split}
\end{equation*}

Now, consider the region $C_{\cap}=C_{s} \cap C_{t_n}$, which is the intersection of the two circles centered at $s$ and $r$ with radii
$\pi_{s} = \min(D_{st_n},\frac{2D_{st_n}}{1+\sqrt[\alpha]{\gamma}})$ and
$\pi_{r} = \min(D_{st_n},\frac{2\sqrt[\alpha]{\gamma}D_{st_n}}{1+\sqrt[\alpha]{\gamma}})$,
respectively. The radii $\pi_s$ and $\pi_r$ takes the bias due to $\alpha$ and $\gamma$ into account. Simply stated, if $\gamma >1$ then
$\pi_{s} < \pi_r$, and if $\gamma <1$ then $\pi_s > \pi_r$, and finally if $\gamma=1$ then the two circles have equal radii. It is clear that
if $\gamma \in (0,\infty)$ then the area of $C_{\cap} >0$.

If the relay is positioned outside $C_{\cap}$, then
\begin{equation*}
\max\left( \frac{2D_{sr}}{1+\sqrt[\alpha]{\gamma}},\frac{2\sqrt[\alpha]{\gamma}D_{rt_n}}{1+\sqrt[\alpha]{\gamma}} \right) > D_{st_n},
\end{equation*}
implying,
\begin{equation*}
f_{sT}(P^{+}_{s},D^{-}_{sT}) > \min \left(f_{sT_1}(P^{+}_{s},D^{-}_{sT_1}),f_{rT_2}(P^{+}_{r},D^{-}_{rT_2}) \right).
\end{equation*}
This means that the min-cut of the path $\{(s,T)\}$ is strictly larger than the min-cut of the path $\{(s,T'_{1}),(r,T'_{2})\}$,
implying that all the multicast flow must be sent
over the path $\{(s,T)\}$, rendering relay useless. Hence, the optimal relay position must lie inside $C_{\cap}$.

From here, it is straightforward to see that if the optimal relay position lies in the interior of $C_{\cap}$, rendering the min-cut of the
path $\{(s,T'_{1}),(r,T'_2)\}$ strictly larger than the path $\{(s,T)\}$; then flow over the path $\{(s,T'_{1}),(r,T'_2)\}$ must be maximized
first and then the flow over the path $\{(s,T)\}$, in order to maximize $R_{sT}$. This proves the first point of the theorem.

Similarly, if the optimal relay position lies on the boundary of the region $C_{\cap}$, then
\begin{equation*}
 f_{sT}(P^{+}_{s},D^{-}_{sT})=\min \left(f_{sT_1}(P^{+}_{s},D^{-}_{sT_1}),f_{rT_2}(P^{+}_{r},D^{-}_{rT_2}) \right),
\end{equation*}
rendering the min-cut of the two paths equal. In this case, all the flow can be sent over the path $\{(s,T)\}$, $\{(s,T_{1}),(r,T_2)\}$ by
arbitrarily sharing the flow between them. This case is reminiscent to the case when relay is placed outside $C_{\cap}$, but for
completeness we count it as an individual case, and moreover in this case relay is not really useless.
This proves the second part and hence completes the proof of
Theorem~\ref{L3}.
\end{IEEEproof}


\section{Proof of optimality of Algorithm ORP}\label{ap:Proof-ORP}
Assume a given $\mathcal{G(N,A)}$ and $\gamma=1$. The argument of the output of Program (A) is a point $l$ and the objective function value of the
output of Program (A) is distance denoted by $D_{Q}$ (where, $Q$ is the set of points of input to Program (A)).

\begin{figure}[tp]
\begin{center}
\psfrag{s}{$s$} \psfrag{1}{$t_{1}$} \psfrag{2}{$t_{2}$} \psfrag{3}{$t_{3}$} \psfrag{a}{$(a)$} \psfrag{b}{$(b)$}
\psfrag{cs}{$C_{s}$} \psfrag{cr}{$C_{r}$} \psfrag{r}{$r$}
\includegraphics[width=1.0\columnwidth]{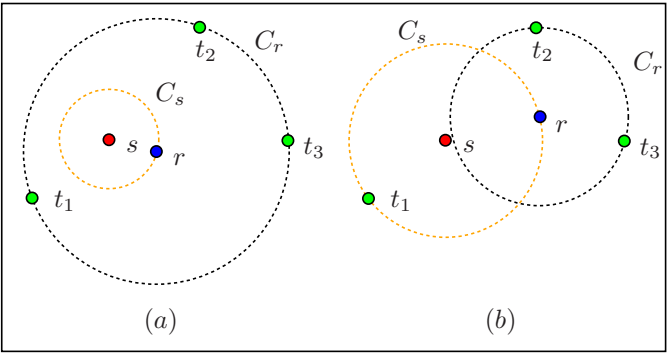}
\end{center}
\vspace{-4mm} \caption{{$T=\{t_{1},t_{2},t_{3}\}$ case illustrating step $3$ of the algorithm ORP.
(a): The relay $r$ is placed at the point $l_{0}=L_{\mathcal{N} \backslash r}(\mathcal{C})$ with $\mathbf{N_{0}}=\{t_{1}\}$. $D_{st_{1}} < D_{rt_{1}}$ and $D_{st_{1}} > D_{sr}$
(b): Reforming
the hyperarc $C_{s}$ and $r$ is placed at $l_{2}=L_{\mathbf{N_{2}}}(\mathcal{C})$ (where, $\mathbf{N_{2}}=\{t_{2},t_{3}\}$), thus reducing
$\max(\pi_{s},\pi_{r})$.
 }} \label{fig:Fig4} \vspace{-6mm}
\end{figure}

\begin{IEEEproof}
In order to prove that Algorithm ORP always outputs optimal relay position, we need to prove that the three steps suffice to
tackle all the topologies of a given $\mathcal{G(N,A)}$ (namely, the distribution of the points $\mathcal{N} \backslash r$ on the $2$-D Euclidean
plane).

First, we divide all the topologies in two classes. In the first, the point $l$ is the optimal relay position (which corresponds to the step $1$
of the algorithm ORP), and the second class in which the point $l$ is not the optimal relay position (this class corresponds to the Step $2$ and $3$
of the algorithm ORP). The only complicated case (if at all) is the Step $3$, so we will go about proving the optimality of the output of
algorithm ORP backwards in the order Step $3$,
then Step $2$ and finally Step $1$.

For a given $\mathcal{G(N,A)}$, compute the point $l_{0}$ defined as,
\begin{equation}
\begin{split}
 l_{0}=&\arg\displaystyle\min_{j \in \mathcal{N} \backslash r} (\max (D_{ij})) \\
&\mbox{subject to:}   \hspace{5mm} i\in \mathcal{C}.
\end{split}
\end{equation}

Form the hyperarcs $C_{s}$ of radius $D_{sl_{0}}$ and $C_{r}$ of radius $D_{\mathcal{N} \backslash r}$.
Denote the value of the quantity $\max(\pi_{s},\pi_{r})=\zeta$.
Construct the set $\mathbf{N_0}=\{t \in T|D_{st} < D_{l_{0}t}  \with D_{l_{0}t} > D_{sl_{0}}\}=\{t'_1,..,t'_{m}\}$, in the increasing order of
distance from $s$. Considering the set $\mathbf{N_{0}}$ is not empty, take the farthest node $t'_{m}$ from $s$. If $D_{st'_{m}} > D_{sl_{0}}$,
then it is clear that $t'_{m}$ should be approached directly from $s$ and not through $r$, because $D_{l_{0}t'_{m}} > D_{sl_{0}}$ and
$\max(D_{sl_{0}}, D_{l_{0}t'_{m}}) < D_{st'_{m}}$. Reforming a source hyperarc $C_{s}$ of radius $D_{st'_{m}}$ (where, $D_{st'_{m}} < \zeta$), the
set $\mathbf{N_{2}}=\{t \in T| D_{st} > D_{st'_m}\}$ can be constructed consisting of all the nodes not lying in the area $C_{s}$. Now, computing
the point $l_{2}=L_{\mathbf{N_{2}}}(\mathcal{C})$ we could form the second hyperarc $C_{r}$ of radius $D_{\mathbf{N_{2}}}$. Note that
$D_{\mathbf{N_{2}}} < \zeta$ because the set $\mathbf{N_{2}}$ consists only the nodes in $T$ that are not in the hyperarc $C_{s}$ and
$D_{st'_m} > D_{sl_{0}}$. Denoting $\max(\pi_{s},\pi_{r})=\zeta''$ (with respect to point $l_{2}$), we now have $\zeta'' < \zeta$. We cannot
further reduce
$\max(\pi_{s},\pi_{r})$ as $t'_{m}$ is the farthest node in $\mathbf{N_{0}}$ that satisfies this property. Thus $l_{2}$ is the optimal relay
position. Figure~\ref{fig:Fig4} illustrates this step for $|T|=3$ case.

On the other hand, if the node $t'_{m}$ satisfies the relation $D_{st'_{m}} \leq D_{sl_{0}}$, it is clear that all the nodes in
$\mathbf{N_{0}}$ could be dropped from the computation of the point $l$ and the set
$\mathbf{N_{1}}=\{\mathcal{N} \backslash (r,\mathbf{N_0})\}$ can be constructed. Therefore, computing the point $l_{1}=L_{\mathbf{N_{1}}}(\mathcal{C})$,
gives the optimal relay position as there is no node in the set $\mathbf{N_{1}}$ that influences the computation of the optimal relay position
unnecessarily. Again,
reforming the hyperarcs and denoting $\max(\pi_{s},\pi_{r})=\zeta'$ (with respect to point $l_{1}$), we can easily see that $\zeta' < \zeta$. The value of $\max(\pi_{s},\pi_{r})$
cannot be reduced further because there is no receiver node in $T$ that is in the set $\mathbf{N_{0}}$ that cannot be encompassed by the area
of union of the two
hyperarcs $C_{s}$ and  $C_{r}$ (constructed with respect to the point $l_{1}$). Thus, in this case $l_{1}$ is the optimal relay position.

Finally, if the set $\mathbf{N_{0}}=\{\emptyset\}$, the point $l_{0}$ is clearly the optimal relay position as there is no receiver node in the
system that is affecting the computation of the relay position and can be dropped off simultaneously. Hence, the three steps of algorithm ORP
always outputs the optimal relay position for a given $\mathcal{G(N,A)}$.

\end{IEEEproof}

The case when $\gamma \neq 1$ is a straightforward generalization and line of argument for the proof of optimality remains the same
for the case of $\gamma = 1$.

\end{document}